\documentclass[prl,reprint,doublecolumn,superscriptaddress, nobalancelastpage,dvipsnames]{revtex4-2}
\usepackage[dvipsnames]{xcolor}

\usepackage{tikz}
\usetikzlibrary{matrix}
\usepackage{tikz-feynman}
\usetikzlibrary{arrows}
\usetikzlibrary{arrows.meta}
\usepackage{pgfplots}
\pgfplotsset{scaled x ticks=false}
\pgfplotsset{scaled y ticks=false}
\usepackage[sort&compress]{natbib}
\usepackage{verbatim}
\usepackage{float}
\usepackage{sidecap}
\sidecaptionvpos{figure}{c}
\usepackage{lipsum}
\usepackage{capt-of}
\usepackage{setspace}
\usepackage{dcolumn}
\usepackage{bm}
\usepackage{graphicx}
\usepackage{amssymb}
\usepackage{amsmath}
\usepackage{mathtools}
\usepackage{booktabs}
\usepackage{physics}
\usepackage{multirow}
\usepackage[a4paper]{geometry}
\newgeometry{top=1.25cm,right=1cm,left=1.25cm,bottom=1.25cm}
\usepackage[colorlinks=true, urlcolor=blue, pdfborder={0 0 0}]{hyperref}
\hypersetup{linkcolor=blue,citecolor=BrickRed}

\DeclareMathOperator*{\argmax}{argmax}
\newcommand{\x}{\boldsymbol{x}}

\draft%
\begin{document}

\title{\Large{\textbf{Reflection Equivariant Quantum Neural Networks\\ for Enhanced Image Classification}}}

\author{Maxwell West} \email{westm2@student.unimelb.edu.au}  \affiliation{School of Physics, The University of Melbourne, Parkville, 3010, VIC, Australia}
\author{Martin Sevior} \affiliation{School of Physics, The University of Melbourne, Parkville, 3010, VIC, Australia}
\author{Muhammad Usman} \email{musman@unimelb.edu.au}  \affiliation{School of Physics, The University of Melbourne, Parkville, 3010, VIC, Australia}
\affiliation{Data61, CSIRO, Clayton, 3168, VIC, Australia}

\begin{abstract}
\noindent
\textbf{Machine learning is among the most widely anticipated use cases for near-term quantum computers, however 
there remain significant theoretical and implementation challenges impeding its scale up.
In particular, there is an emerging body of work which suggests that generic,
data agnostic quantum machine learning (QML) architectures may suffer from severe
trainability issues, with the gradient of typical variational parameters vanishing exponentially in the 
number of qubits. Additionally, the high expressibility of QML
models can lead to overfitting on training data and poor generalisation performance.
A promising strategy to combat both of these difficulties is to construct models which
explicitly respect the symmetries inherent in their data, so-called geometric quantum
machine learning (GQML). In this work, we utilise the techniques of GQML for the
task of image classification, building new QML models which are equivariant with respect to 
reflections of the images. We find that these networks are capable of consistently and
significantly outperforming generic ansatze on complicated real-world image datasets,
bringing high-resolution image classification via quantum computers closer to reality.
Our work highlights a potential  pathway for the future development and implementation 
of powerful QML models which directly exploit the symmetries of data.}
\end{abstract}

\maketitle%

\begin{figure*}[ht]
\begin{center}
    {\includegraphics[width=0.9\textwidth]{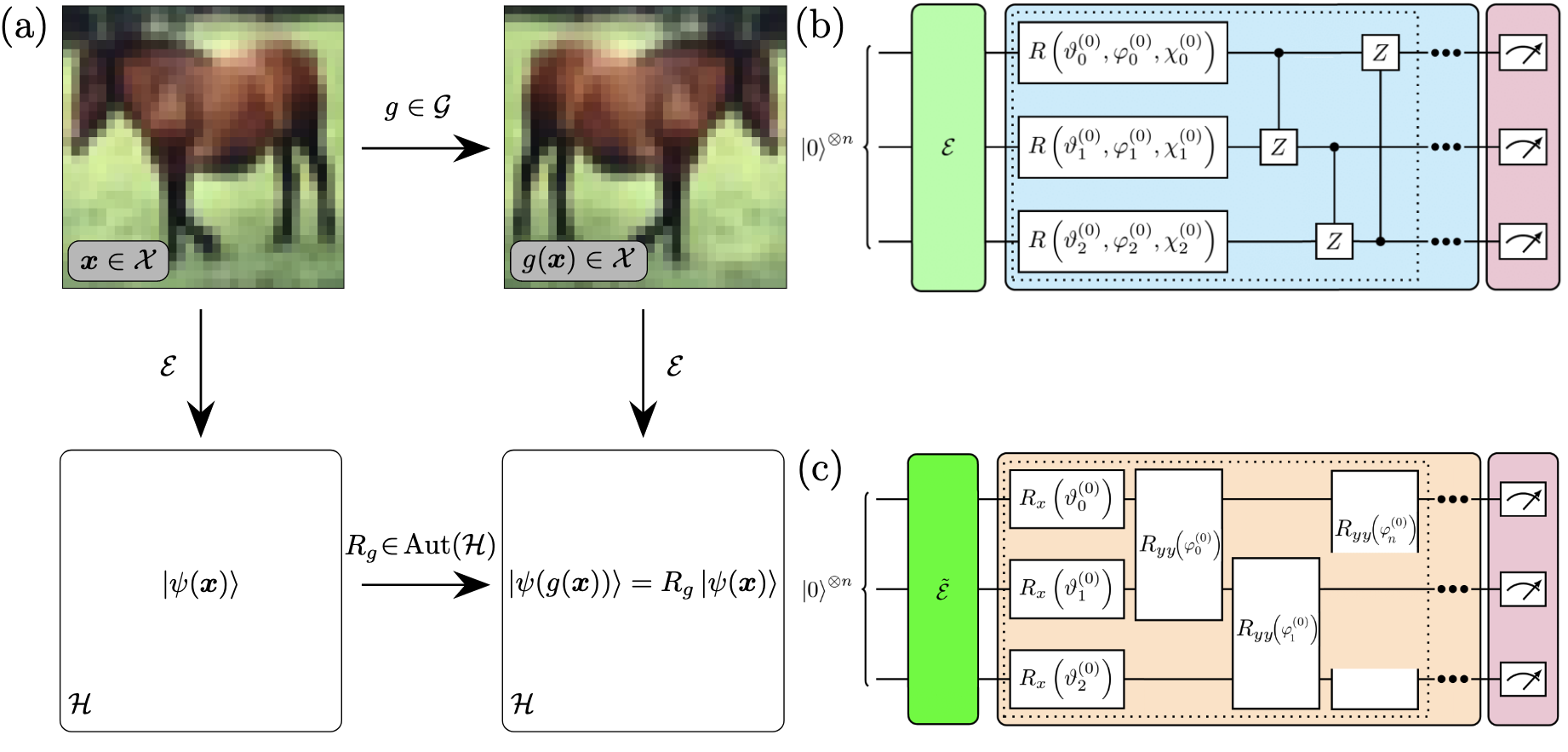}}
  \caption{\label{fig:1}\textbf{Reflection Equivariant Quantum Neural Networks.} (a) We consider image data whose labels are
  left invariant by the action of a group $\mathcal{G}$. In the case shown
  (from the CIFAR-10 dataset~\cite{krizhevsky2009learning}), the label ``horse'' applies to the image both before and after 
  a reflection about the central vertical axis. The action of the symmetry group on the encoded quantum states of the images is 
  determined by a unitary representation $R$ of $\mathcal{G}$ satisfying
  $\ket{\psi(g(\x))}=R_g\ket{\psi(\x)}\ \forall g\in\mathcal{G},\x\in\mathcal{X}$, i.e. the diagram commutes.
  (b) A schematic depiction of the generic quantum variational classifier (QVC) that we employ for image classification.
  The pale green subcircuit $\mathcal{E}$ implements amplitude encoding. The variational component of the circuit consists of
  the unit in the dotted box repeated 100 times (with different parameters in each layer). 
  Finally, $\sigma_z$ measurements are made on each qubit to determine the label prediction.
  (c) A modified circuit which possesses reflection equivariance. The encoding subcircuit $\tilde{\mathcal{E}}$ now includes an additional
  change of basis following the amplitude encoding (see the text and Figure~\ref{fig:amplitudeencoding}). 
  In the variational section, which again consists of the unit in the dotted box repeated 100 times, gates which commute 
  wth the symmetry operations are chosen.
  Finally, $\sigma_z$ measurements are again made to determine the label prediction, although this time products of the measurement outcomes 
  on neighbouring qubits are used in order to retain equivariance (see the text).
  While (b) and (c) depict 3-qubit cartoons of the two models, the actual circuits employ either 10 or 12 qubits depending on 
  the dataset due to the resolution of the images: 1$\times$28$\times$28 for MNIST, and 3$\times$32$\times$32 for CIFAR-10 and CelebA
  (see Figure~\ref{fig:ims}).
  }
\end{center}
\end{figure*}

\twocolumngrid%
\noindent%
\large{{\textbf{1. Introduction}}}
\normalsize
\\ \\
The significant interest in the possibility 
of realising superior machine learning algorithms on quantum computers has led over the 
last few years to intensive study of the capabilities and limitations of quantum machine learning (QML) 
models~\cite{biamonte2017quantum,beer2020training, havlivcek2019supervised, romero2017quantum, dallaire2018quantum,killoran2019continuous,schuld2019quantum,qcnn,schuld2021supervised,tsang2022hybrid}. 
Although remarkable improvements in the capability of QML methods over their classical counterparts have been reported for some specific use cases~\cite{liu2021rigorous,huang2022quantum,west2022benchmarking}, 
the extent to which they can be expected to perform well on general 
datasets remains an open question. In fact, recent work has suggested that generic, commonly employed QML
architectures such as variational quantum circuits will face significant limitations due to their high expressibility and 
potentially low trainability resulting from  barren plateaus in their training 
landscapes~\cite{mcclean2018barren,holmes2022connecting,cerezo2021cost,wang2021noise,pesah2021absence,patti2021entanglement}. 
Efforts to address this, while still maintaining sufficiently complicated circuits to allow for the possibility of quantum advantage, are 
ongoing~\cite{pesah2021absence,patti2021entanglement,skolik2021layerwise,volkoff2021large,grant2019initialization}, but the ultimate resolution of the problem of
barren plateaus remains unclear, and a key research direction
of QML. 
Separately, in the case of image classification, the application of QML is still limited to relatively simple and low resolution datasets~\cite{lu2020quantum}
by the current hardware limitations,
with the development of high performance QML models for complex image datasets another important open problem.
In this work we utilise techniques developed to tackle poor generalisation and barren plateaus to design
new QML models which explicitly exploit the reflection symmetry inherent in many image datasets (see Figure~\ref{fig:1}).
We establish that these reflection equivariant models can provide superior performance on complex images at the
forefront of the current capabilities of QML, both obtaining higher accuracies and possessing parameter gradients
which vanish more slowly than those of a generic counterpart. \\    

Symmetry has long played an important role in physics, facilitating both the discovery and understanding of the laws of nature~\cite{noether}.
More recently, the symmetries of data have been recognised to play an important role in machine learning,
informing the design of some of the most effective classical classification algorithms~\cite{cohen2021equivariant}.
Convolutional neural networks (CNNs), for example, which have been famously successful in the classification of image data,
begin by applying a set of so-called convolutional filters, which act on all regions of the image identically -- they 
have \textit{translation equivariance}~\cite{mnist, kondor2018generalization}. The loss of expressibility that this necessarily entails 
has been found to be more than compensated for by their increased trainability, with CNNs having now dominated image 
classification for years~\cite{mnist, he2016deep}. More generally, in recent years the new field of geometric deep learning 
(GDL)~\cite{gdl, bronstein2021geometric, li2022group,cohen2021equivariant} has begun to explore the role of  symmetry respecting 
neural network architectures beyond such Euclidean translational equivariance, studying for example symmetries of data which lives on  
graphs~\cite{perozzi2014deepwalk} or Riemannian manifolds~\cite{masci2015geodesic}. The observed success of these strategies naturally raises the question of  
whether such techniques could also help to build QML models which benefit from utilising the symmetries of their data.\\

\begin{figure}[h]
 \begin{center}
   \includegraphics[width=0.4\textwidth]{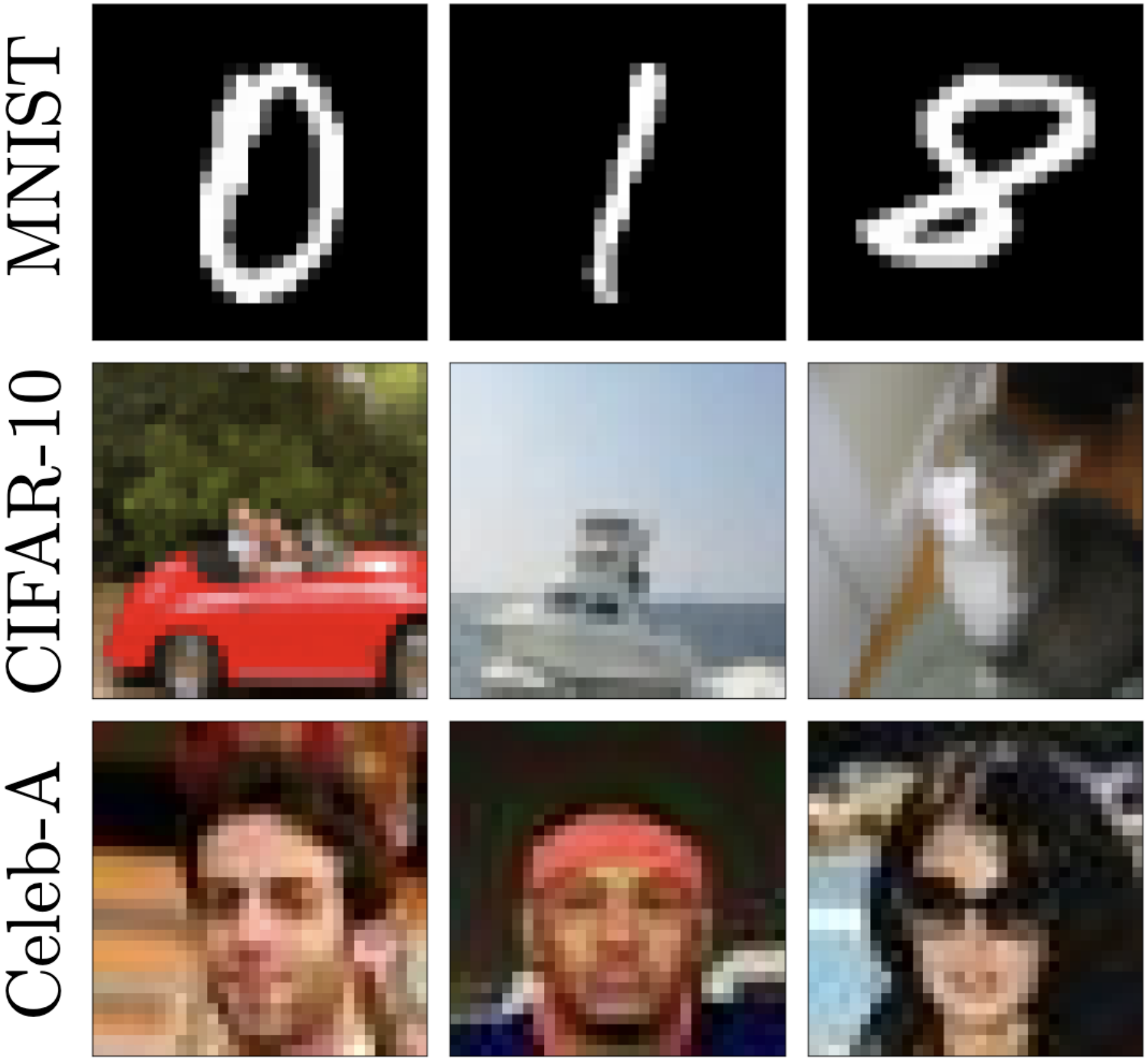}
 \end{center}
  \caption{\label{fig:ims}\textbf{Image datasets.}
  We consider three standard image
  datasets from the ML literature: MNIST~\cite{mnist}, CIFAR-10~\cite{krizhevsky2009learning} and
  CelebA~\cite{yang2015facial}. The MNIST dataset consists of handwritten digits,
CIFAR-10 of various objects and animals, and CelebA of human
faces. In the case of MNIST we restrict our attention to the digits
``0'', ``1'' and ``8''. Having done this, all of the data we consider
has labels which are unchanged under a horizontal reflection and
is therefore suitable for classification by our REQNNs.
  }
\end{figure}

Indeed, promising recent work has incorporated ideas from GDL into QML, resulting in the emerging subject of 
geometric quantum machine learning 
(GQML)~\cite{meyer2022exploiting,ragone2022representation,nguyen2022theory,schatzki2022theoretical,sauvage2022building,skolik2022equivariant,larocca2022group,zheng2022benchmarking}. 
For example, this approach has been used to construct QML models which, classifying data that enjoys 
permutation symmetry, provably avoid barren plateaus~\cite{schatzki2022theoretical}. In this work we turn the techniques of GQML to 
the problem of image classification, an important example of a problem for which deep learning frameworks drastically outperform all other methods.
The translational equivariance of CNNs has already been imported to the quantum setting, with 
quantum convolutional neural networks (QCNNs) offering a pathway to efficient image classification on quantum computers~\cite{qcnn}. 
Here we consider the alternative symmetry of reflections, noting that the labels assigned to an image will often be independent of 
reflections of that image about various axes (see Figure~\ref{fig:1}(a)), as for example in most common object identification problems. 
We construct reflection equivariant quantum neural networks (REQNNs, see Figure~\ref{fig:1}(c)) and show that they  consistently outperform 
a generic model (see Figure~\ref{fig:1}(b)) when benchmarked across three standard image datasets (CIFAR-10 \cite{krizhevsky2009learning}, 
MNIST \cite{mnist}, and Celeb-A \cite{yang2015facial} -- see Figure~\ref{fig:ims} for examples of images from each dataset),
despite having access to fewer trainable parameters. 
These results provide concrete evidence supporting
the emerging philosophy that sacrificing generality and expressibility in favour of targeting 
a smaller but more meaningful fraction of the model space is a promising approach forward for QML.
Moreover, by demonstrating enhanced performance for image classification
through the use of reflection equivariant networks, separate to the previous use of a translation equivariant convolutional
structure in QCNNs~\cite{qcnn}, our results encourage the future development of QML models which exploit as much of the
available symmetry information of their data as possible, including extensions to additional symmetries and the
simultaneous consideration of multiple symmetries. The practical realisation of tailored QML models such as these
will bring the possibility of successfully applying quantum
computing to the many domains of ML, both image-based and otherwise, closer to reality.\\

\noindent
\large{{\textbf{2. Geometric Quantum Machine Learning}}} 
\normalsize
\\ \\ 
We begin by briefly summarising the aspects of GQML relevant to our construction of REQNNs, introducing the notation and formalism required to discuss equivariance and symmetry operations on data encoded into quantum states. 
Interested readers can find further details of GQML in Refs. \cite{ragone2022representation,nguyen2022theory}.
We consider the classification of (image) data $\x\in \mathcal{X}$, with associated labels $y(\x)\in\mathcal{Y}$. 
Our QML models will follow a standard three-step procedure: a data encoding circuit which maps the classical image data $\x$ to a 
quantum state, $\x\mapsto \ket{\psi(\x)}$, followed by a variational circuit $U_{\theta}$, followed by measurements of a set 
of operators $\{M_j\}_{j=1}^{n_\mathrm{classes}}$ to determine the class label. 
For a given set of parameters $\theta\in\Theta$
the prediction $\hat{y}_{\theta}(\x)$ of the model on an input $\x$ is given by
\begin{equation}
  \hat{y}_{\theta}(\x) = \argmax_j \bra{\psi(\x)}\mathcal{U}_{\theta}^\dagger M_j \mathcal{U}_{\theta}\ket{\psi(\x)}\label{eq:prediction}
\end{equation}
We also introduce a symmetry group $\mathcal{G}$ which we will identify with its action on $\mathcal{X}$, i.e. writing
$\mathcal{G}:\mathcal{X}\to\mathcal{X}$, with $g(\x)$ for $g\in\mathcal{G}$ the image obtained by performing the symmetry 
transformation $g$ on $\x$. We emphasise that the symmetries are expected to be respected only at the level of the labels of the data, i.e.
$y(g(\x))=y(\x)\ \forall \x\in\mathcal{X},\ g\in\mathcal{G}$, but perhaps $g(\x)\neq\x$. 
In this work we will focus on the group which enacts reflections about the central vertical axis of the images, 
so that $\mathcal{G}\cong\mathbb{Z}_2$.
The group will act on the Hilbert space $\mathcal{H}$ of the quantum computer via a unitary 
representation $R : \mathcal{G}\to \mathrm{Aut}(\mathcal{H})$ which satisfies
$R(g)\ket{\psi(\x)}=\ket{\psi(g(\x))}$. 
Henceforth we write $R_g\equiv R(g)$.
We wish for the predictions of our QNN to be $\mathcal{G}$-invariant, i.e.
\begin{equation}
  \hat{y}_{\theta}(\x) = \hat{y}_{\theta}(g(\x)) \qquad \forall g\in\mathcal{G},\x\in\mathcal{X}, \theta\in\Theta \label{eq:invariant} 
\end{equation}
\noindent
From Equation~\ref{eq:prediction} we have that

\begin{align}
  \hat{y}_{\theta}(g(\x)) &= \argmax_j \bra{\psi(g(\x))}\mathcal{U}_{\theta}^\dagger M_j \mathcal{U}_{\theta}\ket{\psi(g(\x))}\\
  &=\argmax_j \bra{\psi(\x)}R_g^{\dagger}\mathcal{U}_{\theta}^\dagger M_j \mathcal{U}_{\theta}R_g\ket{\psi(\x)}
\end{align}

\noindent
and therefore the condition of Equation~\ref{eq:invariant} will be satisfied if 
\begin{equation}
  \left[ R_g,\ \mathcal{U}_{\theta}^\dagger M_j \mathcal{U}_{\theta} \right] = 0\qquad \forall g\in\mathcal{G},\x\in\mathcal{X}, \theta\in\Theta \label{eq:equivariant} 
\end{equation}

\noindent
We refer to QNNs satisfying this condition as reflection equivariant.\\

\noindent
\large{{\textbf{3. Results}}} 
\normalsize
\\ \\ 
Our first task in building REQNNs is to establish the unitary representation of the symmetry group $\mathcal{G}$.
As in our case $\mathcal{G}\cong\mathbb{Z}_2$, there is only one nontrivial symmetry operator $R_g$, namely the one which maps 
a state to the state encoding the reflection of the image encoded by the original state.
As we have $R_g \ket{\psi(\x)} = \ket{\psi(g(\x))}$ for all images $\x$, $R_g$ will be
determined by the form of the data encoding map, to which we now turn. Due to our desire to classify high dimensional
image data using only the relatively small number of qubits available to classical simulators we need to employ a method
of encoding which is highly efficient in the number of qubits required. For this reason we choose amplitude encoding, i.e.,
if $\x$ is a vector containing the pixel values of an image then we construct the state
\[ \x\mapsto\ket{\psi(\x)} =\sum_i x_i\ket{i}  \]
As there are $2^n$ amplitudes available for an $n$ qubit state,
only $\lceil \log_2 (C × L × W )\rceil$ qubits are needed, where $C$ is the
number of channels of the image, $L$ is the length, and $W$
the width. For MNIST, we have $(C, L, W ) = (1, 28, 28)$ and
therefore require 10 qubits, and for CIFAR-10 and CelebA
$(C, L, W ) = (3, 32, 32)$, requiring 12. In both cases we append
and prepend zeros equally in order to obtain an input vector of
length a power of two in order to proceed with the amplitude
encoding.\\

\begin{figure}[h]
 \begin{center}
   \includegraphics[width=0.49\textwidth]{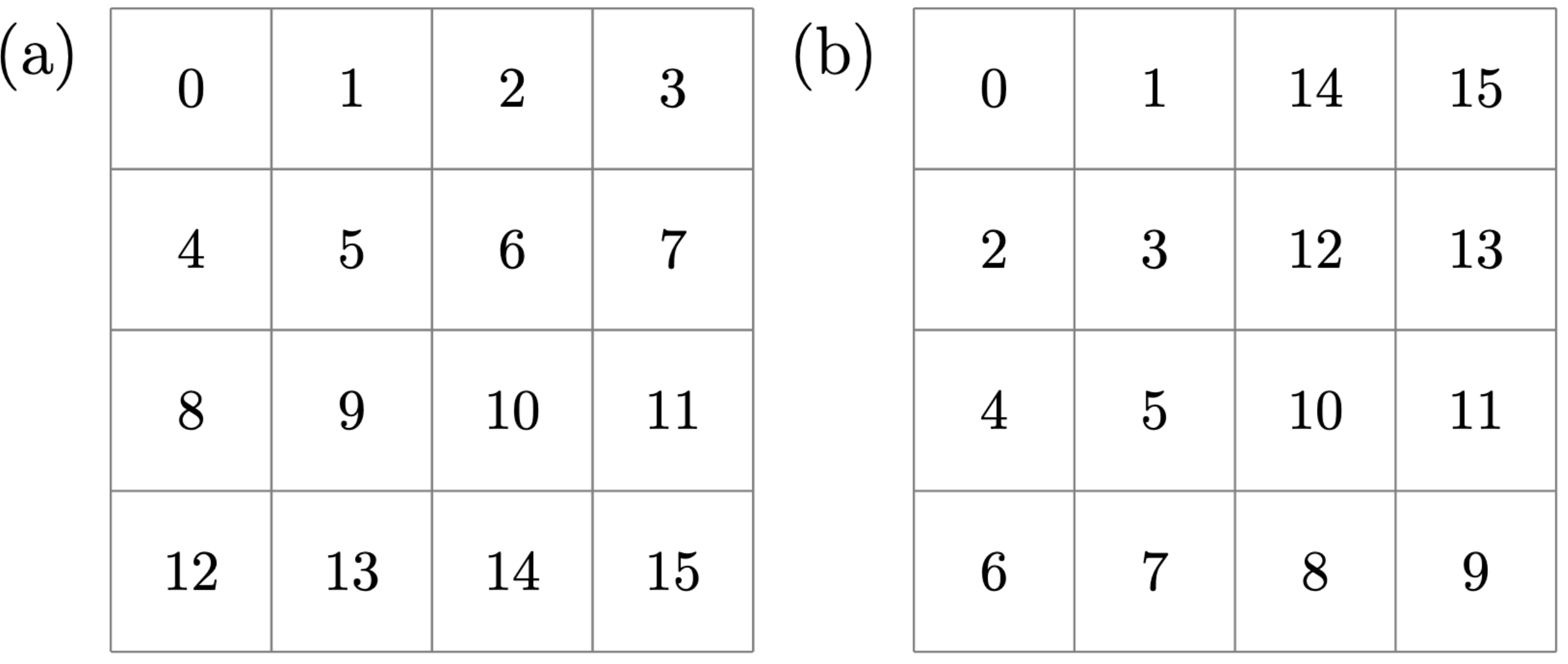}
 \end{center}
  \caption{\label{fig:amplitudeencoding}\textbf{Amplitude Encoding.} 
  Encoding for an example $4\times 4$ image. The pixels are numbered by the index of the basis 
  state into the amplitude of which they are encoded. (a) The standard case. For example, the pixel value of
the pixel in the top right corner is encoded into the amplitude of
  $\ket{3}\equiv \ket{0011}$. (b) Here we permute the order of the encoding. For
example, the pixel value of the pixel in the top right corner is now
  encoded into the amplitude of $\ket{15}\equiv \ket{1111}$. In this case a horizontal reflection is represented by $X^{\otimes n}$.
  }
\end{figure}

\begin{figure*}[ht]
 \begin{center}
   \includegraphics[width=0.75\textwidth]{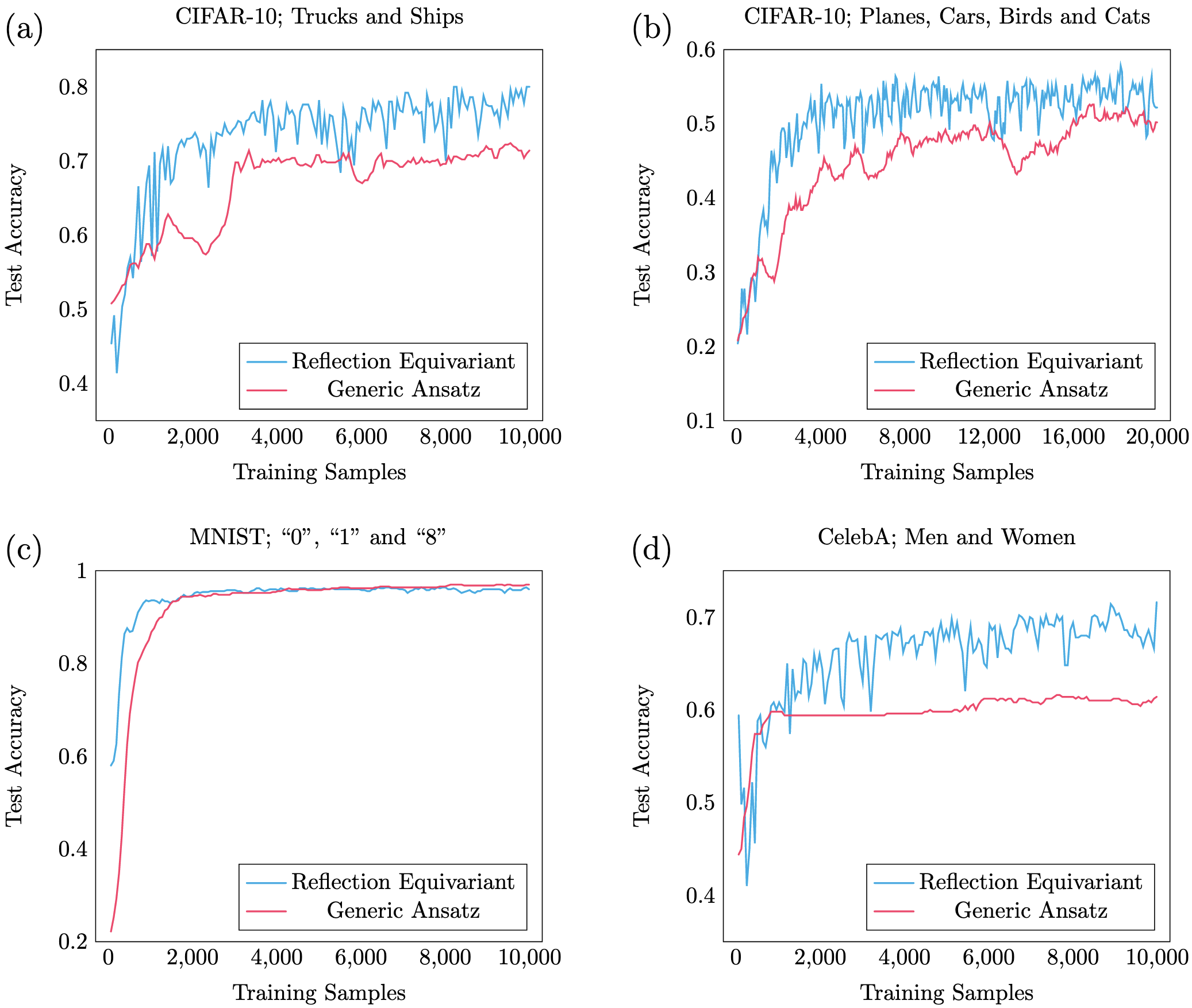}
  \caption{\label{fig:results}\textbf{Quantum neural network performances. }
   We compare the performance of the generic and reflection equivariant QNNs across a range of image datasets.
   As the different datasets have varying resolutions, the number of qubits needed to implement the models
   also changes (see the Results section), providing another axis along which to contrast the performance of the two classes of models.
   We find that the reflection equivariant QNNs learn more quickly than their generic counterparts consistently across different datasets,
   number of classes, and number of qubits used to implement the QNNs.
   The plotted accuracies refer to 500 test samples, calculated at various
points throughout the training process. 
   (a, b) Two and four class classification using the CIFAR-10 dataset, respectively. 
   (c) Three class classification with the MNIST dataset. Although this dataset is quite simple, and QNNs have previously been used to achieve
   high accuracy on all ten classes~\cite{west2022benchmarking}, we restrict here to the digits ``0'', ``1'' and ``8'' as they are the only ones which (approximately)
   respect the reflection symmetry. Both models achieve high test accuracy ($>$96\%), but the reflection equivariant model learns more quickly.
   (d) The CelebA dataset consists of images of human faces. We consider the classification task of determining the gender of the imaged person,
   again finding that the reflection equivariant model significantly outperforms its generic counterpart.
   }
 \end{center}
\end{figure*}

Unfortunately, standard amplitude encoding will render
$R_g$ a complicated, non-local operator which will be difficult
to work with in practice, especially on real hardware with
constrained qubit connectivity. In order to rectify this we
consider, for our equivariant models, a slight modification of
standard amplitude encoding which entails rearranging the
order in which the pixel values are encoded so as to produce
an encoding with respect to which we have $R_g = X^{\otimes n}$ (see
Figure~\ref{fig:amplitudeencoding}). The standard order in which the pixel values
are read is shown in Figure~\ref{fig:amplitudeencoding}(a) for an example of a 4 × 4
greyscale (i.e. only one channel) image, and our alternate
encoding in Figure~\ref{fig:amplitudeencoding}(b). The encoding strategy of Figure~\ref{fig:amplitudeencoding}(a)
is employed in the generic model (Figure~\ref{fig:1}(b)), and the strategy of Figure~\ref{fig:amplitudeencoding}(b) in the equivariant
model (Figure~\ref{fig:1}(c) of the main text). In the second case, a
reflection of the image is represented at the Hilbert space
level by the operator $X^{\otimes n}$, which exchanges the basis states $\ket{i}$ 
and $\ket{2n-1-i}$. The case of three channel RGB images
is similar, with the order that the data is encoded chosen so
as to enforce the requirement that, for every pixel and every
channel, the amplitude of the states $\ket{i}$ and $\ket{2n-1-i}$ are
the values (in a given channel) of a pair of pixels related
by the vertical reflection. With this choice of encoding we have that
$R_g \ket{\psi(\x)} = \ket{\psi(g(\x))}$ for $g=e, r$ the identity and horizontal
reflection operations on the images respectively, with $R_e = I$,
$R_r = X^{\otimes n}$. Armed with the representation of our symmetry
group (consisting simply of the operators $\{I, X^{\otimes n}\}$) we are
ready to begin constructing REQNNs.\\

Given a symmetry group $\mathcal{G}$ and a unitary representation $R$, various ways of constructing  equivariant QNNs have been proposed.
Ref. \cite{meyer2022exploiting}, for example, takes a standard set of gates and ``symmetrises'' them, 
building new gates which are guaranteed to commute with the symmetry representation.
Here, benefiting from the simplicity of our representation,  we adopt a different approach, 
simply manually selecting gates and measurements which commute with $X^{\otimes n}$
and therefore satisfy Equation~\ref{eq:equivariant}.
Schematic depictions of the two models considered in this work are shown in Figure~\ref{fig:1}(b,c).
The model of Figure~\ref{fig:1}(b) is a ``generic ansatz'' consisting of amplitude encoding followed by a standard
variational circuit followed by $\sigma_z$ measurements on the first $n_{\mathrm{classes}}$ qubits. The prediction
of the model is taken to be the class corresponding to the qubit which reports the largest such measurement 
(i.e. $M_j=\sigma^{(j)}_z$ in the notation of Equation~\ref{eq:prediction}, with $\sigma^{(j)}_z$ the Pauli $z$
operator acting on the $j$th qubit). The reflection equivariant model of Figure~\ref{fig:1}(c) differs in several ways.
First, the encoding stage is slightly modified as previously discussed (see also Figure~\ref{fig:amplitudeencoding}). Second, 
the variational component is built from $R_x$ and $R_{yy}$ gates, both of which commute with $X^{\otimes n}$.
Finally, the class labels are determined by measurements $M_j=\sigma^{(j)}_z \otimes \sigma^{(j+1\ \mathrm{mod}\ n)}_z$,
which also commute with $X^{\otimes n}$. This QNN therefore satisfies the equivariance condition of Equation~\ref{eq:equivariant}.\\

We implement the networks within the Pennylane framework~\cite{bergholm2018pennylane}, and train them using the Nesterov momentum 
optimiser~\cite{nesterov1983AMF}.
The results are shown in Figure~\ref{fig:results}. 
Our results show that the reflection equivariant model consistently outperforms the generic model, despite
its lower expressibility and the generic model containing 50\% more trainable parameters.
The increased performance is particularly notable in the cases of the more complicated datasets, CIFAR-10 and Celeb-A,
which consist of colour (RGB) images. In the case of MNIST, and although the difference in final test accuracies is 
negligible, we nonetheless see the reflection equivariant model learning more quickly in the initial stage of training, 
with its focus narrowed to the more meaningful subset of reflection insensitive decision functions.\\

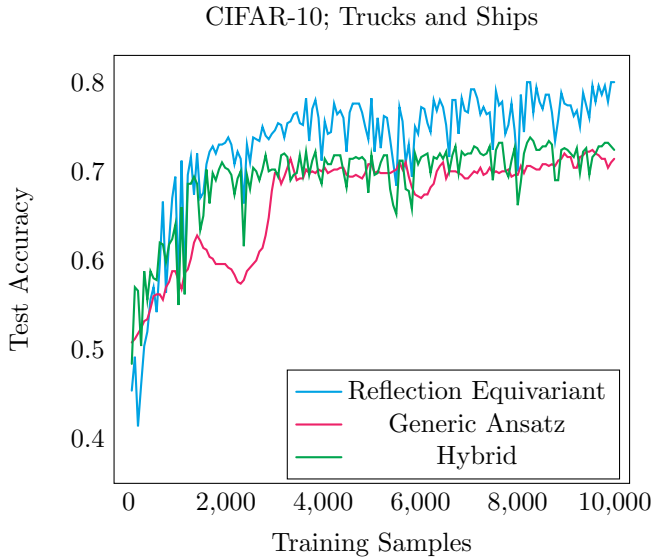
\begin{figure}[t]
 \begin{center}
 \begin{tikzpicture}[line cap=round]
  \begin{axis}[
      xmin=-300, xmax=10300,
      ymin=0.35, ymax=0.83,
      ytick style={draw=none},
      xtick style={draw=none},
      xlabel=Training Samples,
      ylabel=Test Accuracy,
      title={CIFAR-10; Trucks and Ships},
      xtick={0,2000, 4000, 6000, 8000, 10000},
    legend style={at={(0.335,0.1375)},anchor=west, cells={align=left}}
    ]
    \addplot[Cerulean, thick] table {symcifar100_output};
    \addplot[WildStrawberry, thick] table {cifar100_output};
    \addplot[Green, thick] table {norea_symcifar100_output};
    \legend{Reflection Equivariant, Generic Ansatz, Hybrid};
  \end{axis}

\end{tikzpicture}
 \end{center}
  \caption{\label{fig:no_rea}\textbf{Effect of reflection equivariance.} 
  In addition to the considered generic and reflection equivariant models we trial a ``hybrid'' model with the 
  same variational circuit structure and final measurement strategy
  as the reflection equivariant model, but which uses the standard amplitude encoding map ${\mathcal{E}}$ 
  (Figure~\ref{fig:amplitudeencoding} (a))
  instead of the modified map $\tilde{\mathcal{E}}$ (Figure~\ref{fig:amplitudeencoding} (b)).
  Therefore, the hybrid model continues to commute with $X^{\otimes n}$,
  but this operator no longer represents a meaningful transformation of the data.
  The considerable drop in accuracy from the reflection equivariant model, 
  despite sharing the same variational architecture, confirms the importance of respecting meaningful symmetries of the data.
  }
\end{figure}

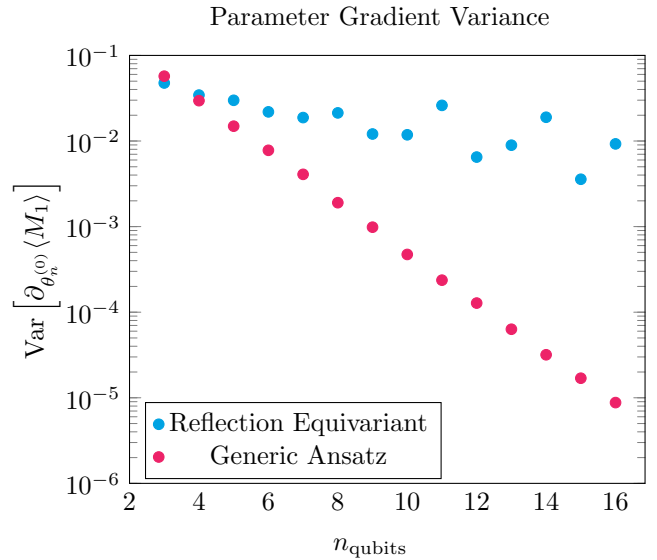
\begin{figure}[h]
 \begin{center}
 \begin{tikzpicture}[line cap=round]
  \begin{semilogyaxis}[
      xmin=2, xmax=16.85,
      ymin=0.000001, ymax=0.1,
      title={Parameter Gradient Variance\phantom{p}},
      xtick style={draw=none},
      xlabel=$n_{\mathrm{qubits}}$\phantom{Tp},
      ylabel={$\mathrm{Var}\left[ \partial_{\theta_n^{(0)}}\langle M_1 \rangle \right]$},
    legend style={at={(0.03,0.1)},anchor=west, cells={align=left}}
    ]
    \addplot[only marks, Cerulean,skip coords between index={0}{14}, scatter,scatter/classes={a={Cerulean}}] table {var};
    \addplot[only marks, WildStrawberry,skip coords between index={14}{28}, scatter,scatter/classes={b={WildStrawberry}}] table {var};
    \legend{Reflection Equivariant, Generic Ansatz};
  \end{semilogyaxis}

\end{tikzpicture}
 \end{center}
  \caption{\label{fig:var}\textbf{Variance of parameter gradients.} 
  We calculate the
derivative of the expectation value of the measured operator $M_1$ in
both the generic and reflection equivariant cases with respect to the
first variational parameter in the final layer of the circuits. This is
repeated for 3000 random circuit initialisations, and the variance of
the results is plotted as a function of the number of qubits. At the
  number of qubits relevant for this work ($n_{\mathrm{qubits}} = 10, 12$) we find
that the variance of the gradients in the equivariant case is several
orders of magnitude higher than the generic case. This is consistent with the improved
trainability observed for the equivariant networks, as well as the
  increased volatility of the test accuracies seen in Figures~\ref{fig:results} and~\ref{fig:no_rea}.}
\end{figure}

As the reflection equivariant and generic classifiers considered in this work are constructed from substantially different quantum circuits,
there is a possibility that the enhanced performance of the reflection equivariant model is due to some other factor than its respecting 
of the symmetry.
In order to separate the effect of simply moving to a different circuit architecture from the role played by the reflection equivariance
we consider a ``hybrid'' model consisting of the standard encoding $\mathcal{E}$ of Figure~\ref{fig:1}(b) (see also Figure~\ref{fig:amplitudeencoding}(a)),
and the variational body and measurements of Figure~\ref{fig:1}(c). This produces a model with the same restricted expressibility
of the reflection equivariant model, which remains equivariant with respect to the operator $X^{\otimes n}$,
but for which $X^{\otimes n}$ no longer enacts a meaningful symmetry.
The results, shown in Figure~\ref{fig:no_rea}, show the reflection equivariant model significantly outperforming
this hybrid model, confirming the importance of building networks which respect genuine symmetries of the data.\\

Being represented by a group with only two elements, we do
not expect the respecting of reflection symmetry to lead to a
provable avoidance of barren plateaus as has been previously
reported in the case of permutation symmetry~\cite{schatzki2022theoretical}, but our
results here are an encouraging sign that, even in the absence
of such guarantees, considerable gains in accuracy may be
realised by such models in practice. To explore this further
we numerically investigate the variance of the gradient of a
measured Pauli observable in both the generic and reflection
equivariant cases (see Figure~\ref{fig:var}, c.f. Figure 3 of Ref.~\cite{mcclean2018barren}).
As expected, the (universal) generic model experiences a
barren plateau, with exponentially vanishing gradients.
While we also see approximately exponential decreases in
the (non-universal) reflection equivariant model, the rate of
decrease is drastically reduced, indicating that training will
remain feasible for much larger quantum circuits. This is
consistent with our expectation that, while the REQNNs
may also asymptotically experience barren plateaus, they
will be able to offer enhanced performance in the highly
interesting region of several tens of qubits explored in this
work and accessible on near-term hardware.
\\ 

\noindent
\large{{\textbf{4. Conclusion}}} 
\normalsize
\\ \\
Geometric quantum machine learning is rapidly emerging as a promising research direction which 
may ameliorate two of the key challenges facing QML: barren plateaus and overfitting.  
We have applied the techniques of GQML to the task of creating QML models
for image classification which are equivariant with respect to horizontal reflections, finding a consistent improvement over 
generic, symmetry agnostic models, despite those models being more expressive. 
These encouraging results join the previous GQML literature~\cite{meyer2022exploiting,ragone2022representation,nguyen2022theory,schatzki2022theoretical,sauvage2022building,skolik2022equivariant,larocca2022group,zheng2022benchmarking} 
in demonstrating clear advantages in building tailored
QML models which respect the symmetry of the data they are attempting to classify, rather than simply employing
universal models, and provide further evidence of the potential of this research direction in the NISQ era and beyond. 
\\


The authors acknowledge useful discussions with Jamie
Heredge. MW acknowledges the support of the Australian Government Research Training Program Scholarship. 
Computational resources were provided by the National Computing Infrastructure (NCI) and Pawsey Supercomputing Centre through the 
National Computational Merit Allocation Scheme (NCMAS). This work was supported by Australian Research Council Discovery Project DP210102831.

\def\bibsection{\subsection*{\refname}}

\bibliographystyle{IEEEtran}
\bibliography{./refs}

\end{document}